\documentclass[a4paper,UKenglish,cleveref, autoref, thm-restate]{lipics-v2021}

\usepackage{fontawesome5}
\usepackage[most]{tcolorbox}
\usepackage{float}
\usepackage{booktabs}
\usepackage{pgfplots}
\pgfplotsset{compat=1.18}
\usepackage{subcaption}
\usepackage{tikz}
\newcommand{\revision}[1]{{\color{black}#1\normalfont}}
\usepackage{array}

\newcolumntype{L}[1]{>{\raggedright\arraybackslash}m{#1}}
\newcolumntype{R}[1]{>{\raggedleft\arraybackslash}m{#1}}

\newcommand{\hbarrow}[4]{%
{\fontsize{6}{6.5}\selectfont #1} &
\begin{tikzpicture}[baseline=-0.5ex]
    \fill[gray!70] (0,0) rectangle ({#2/#3*#4},0.08);
    \node[anchor=west,font=\fontsize{6}{6.5}\selectfont]
        at ({#2/#3*#4+0.04},0.04) {#2};
\end{tikzpicture}
\\[-0.9mm]
}

\newcommand{\hbarrowpct}[3]{%
\hbarrow{#1}{#2}{#3}{2.7}
}

\newcommand{\hbarrowpctsmall}[3]{%
\hbarrow{#1}{#2}{#3}{2.0}
}

\newcommand{\likertbarcompactlr}[5]{%
\begin{minipage}[t]{0.315\textwidth}
\centering
\scriptsize
\textbf{#1} #2

\vspace{1mm}

\begin{tikzpicture}
\begin{axis}[
    ybar,
    width=\linewidth,
    height=2.6cm,
    ymin=0,
    ymax=75,
    bar width=6pt,
    axis x line*=bottom,
    axis y line=none,
    symbolic x coords={1,2,3,4,5},
    xtick=data,
    xticklabels={1,2,3,4,5},
    x tick label style={font=\scriptsize},
    nodes near coords={
        \pgfmathprintnumber[fixed,precision=1,zerofill=false]{\pgfplotspointmeta}\%
    },
    nodes near coords style={font=\tiny, yshift=1pt},
    tick style={draw=none},
    enlarge x limits=0.15,
]
\addplot[fill=gray!70, draw=none] coordinates {#5};
\end{axis}
\end{tikzpicture}

\vspace{0.5mm}

\parbox[t]{0.43\linewidth}{\raggedright\tiny #3}%
\hfill
\parbox[t]{0.43\linewidth}{\raggedleft\tiny #4}
\end{minipage}%
}

\usepackage{listings}
\newcommand{\participants}{66}
\newcommand{\participantsScrum}{33}
\newcommand{\participantsRefactoring}{33}
\newcommand{\country}{\textit{Brazil}} 

\title{\revision{AI-Conducted Interviews in Empirical Software Engineering: An Experience Report}} 

\author{Rohit Gheyi}{Federal University of Campina Grande}{rohit@dsc.ufcg.edu.br}{https://orcid.org/0000-0002-5562-4449}{}

\author{Danyllo Albuquerque}{VIRTUS, Federal University of Campina Grande}{danyllo.albuquerque@virtus.ufcg.edu.br}{https://orcid.org/0000-0001-5515-7812}{}

\author{Márcio Ribeiro}{Federal University of Alagoas}{marcio@ic.ufal.br}{https://orcid.org/0000-0002-4293-4261}{}

\author{Mirko Perkusich}{VIRTUS, Federal University of Campina Grande}{mirko@virtus.ufcg.edu.br}{https://orcid.org/0000-0002-9433-4962}{}

\authorrunning{Gheyi et al.} 

\Copyright{Gheyi et al.} 



\category{} 

\relatedversion{}
\ArticleNo{1}

\begin{document}
\nolinenumbers

\maketitle

\begin{abstract}
\textbf{Background:}
Semi-structured interviews are widely used in empirical software engineering (ESE) to understand practitioners' perceptions, practices, challenges, and experiences. \revision{However, interviews are resource-intensive, especially when participants have different schedules, locations, and preferred natural languages.} \revision{Recent advances in large language models (LLMs) create opportunities for interviews conducted by artificial intelligence (AI) systems, but evidence about their operational use and methodological limitations in software engineering studies remains limited.}
\textbf{Aims:}
\revision{This experience report examines a customized MyGPT used to conduct short, self-administered semi-structured interviews in two ESE studies: one on refactoring practices and another on the use of generative AI in Scrum-related activities.} \revision{We characterize operational feasibility among valid submissions, participant perceptions, practical benefits and limitations, and lessons for researchers; we do not evaluate equivalence with human-conducted interviews or the fidelity of the generated summaries to full transcripts.}
\textbf{Method:}
We designed two AI-based interview protocols and deployed them through shareable links. \revision{Participants accessed the AI interviewer on their own devices, used voice interaction, selected or requested a preferred natural language, and completed a short interview without a researcher present.} \revision{The AI interviewer followed a predefined protocol and generated a structured synthesis that participants voluntarily submitted; these AI-mediated artifacts were not treated as verbatim transcripts.} \revision{Over a three-day period, we analyzed \participants{} submissions across the two studies and a post-interview questionnaire about clarity, comfort, naturalness, willingness to participate again, and related perceptions. We also performed a post-hoc structural audit of artifact format, predominant natural language, length, and consistency with the protocol reported in the questionnaire.}
\textbf{Results:}
\revision{Among the analyzed respondents, the workflow was completed and generally well perceived. A post-hoc audit found that 92.4\% (61/66) of the submitted artifacts followed the expected structured-synthesis format, that 65/66 artifacts were written predominantly in Portuguese and one in English, and that two submissions (3.0\%, 2/66) contained refactoring content although the corresponding questionnaire record identified the Scrum protocol. Because interaction language was not recorded as a structured variable, artifact language is used only as a proxy.} \revision{Overall experience was rated positively by 90.9\% (60/66), 90.9\% (60/66) felt comfortable, 95.5\% (63/66) considered the questions clear, 97.0\% (64/66) rated the pace positively, and 89.4\% (59/66) would participate again.} \revision{Because the workflow was asynchronous, successful completers did not need to schedule a synchronous session with a researcher, and the procedure produced an immediately available structured artifact; however, we did not measure completion rates, researcher time savings, or summary fidelity.} \revision{The comparative questionnaire item produced mixed responses: 42.4\% (28/66) selected its midpoint, 37.9\% (25/66) leaned toward the human-interviewer endpoint, and 19.7\% (13/66) leaned toward the AI-interviewer endpoint.} \revision{The most frequent limitations were generic questions and limited sensitivity to answers, each selected by 28.8\% (19/66), followed by insufficient depth at 27.3\% (18/66), and privacy concerns and missed human interaction, each at 22.7\% (15/66).}
\textbf{Conclusions:}
\revision{The findings support the operational viability and acceptability of this MyGPT-based workflow among participants who completed and submitted the procedure.} \revision{They do not establish that AI-conducted interviews produce data comparable in richness, accuracy, or analytical value to human-conducted interviews.} \revision{AI interviewers should therefore be treated as a complementary option for short, focused, low-risk workflows, with careful protocol design, transparent platform-related privacy information, validation of AI-generated artifacts, and human oversight.}
\end{abstract}

\section{Introduction}
\label{sec:introduction}

Semi-structured interviews are a central method in empirical software engineering (ESE)~\cite{seaman1999qualitative,hove2005experiences}. They allow researchers to investigate how software practitioners and students perceive, adopt, adapt, and reason about tools, processes, and development practices in real-world settings. Interviews are especially valuable when the phenomenon under study is emerging, context-dependent, or insufficiently captured by quantitative instruments alone.
However, conducting interviews is resource-intensive~\cite{hove2005experiences}. Researchers need to recruit participants, schedule sessions, conduct interviews, ask follow-up questions, transcribe conversations, and organize qualitative data. \revision{These challenges become more pronounced when participants are geographically distributed, have limited availability, or prefer different natural languages.} Interviews also require a balance between consistency and flexibility: the protocol must support comparison across participants while still allowing clarification, probing, and contextualized answers.

Recent advances in generative artificial intelligence \revision{(AI)}, largely enabled by the Transformer architecture~\cite{vaswani2017attention} and later demonstrated at scale in large language models \revision{(LLMs)}~\cite{brown2020language}, create new opportunities for supporting qualitative data collection~\cite{chopra2023conducting,cuevas2023qualitative}. LLMs can interact conversationally, ask follow-up questions, adapt to different languages, and produce structured summaries. These capabilities suggest that AI-based interviewers may reduce some operational barriers of interview studies by allowing participants to complete interviews independently, through a shareable link, using voice interaction, and without scheduling a synchronous session with a human interviewer.
At the same time, AI-conducted interviews raise important methodological questions~\cite{chopra2023conducting,cuevas2023qualitative,zhang2025harnessing}. An AI interviewer may ask shallow follow-up questions, misunderstand participant responses, handle language changes inconsistently, or generate summaries that omit relevant details. Participants may also react differently to the absence of a human interviewer: some may appreciate the convenience and perceived neutrality of speaking to AI, whereas others may find the interaction unusual, less natural, or less engaging. Therefore, before AI-conducted interviews can be adopted more broadly in ESE, we need concrete evidence on how they work in practice, how participants perceive them, and what limitations researchers should anticipate.

In this article, we report our experience using an AI interviewer, implemented as a customized conversational agent, to conduct semi-structured interviews in two ESE studies. The first study investigated how software professionals perform refactorings~\cite{Opdyke-SOOPPA-1990, Fowler-book-1999} in real projects, including refactoring frequency, tool use, quality assessment, and behavior preservation. The second study investigated how professionals and students use generative AI in Scrum-related activities~\cite{schwaber2001agile,masood2020real}, including backlog refinement, user story writing, acceptance criteria definition, documentation, planning, and decision support.
For both studies, we designed AI-based interview protocols and deployed them through shareable links. \revision{Participants accessed the AI interviewer using their own device, used voice interaction, selected or requested a preferred natural language, and completed a short semi-structured interview without the presence of a researcher.} \revision{At the end, participants were instructed to request a structured synthesis and voluntarily submit that AI-mediated artifact to the research team.} \revision{Because the research team did not systematically collect the full conversations, the submitted artifacts cannot be assumed to preserve participants' wording, detail, or conversational context.} They also answered a post-interview questionnaire about their experience of being interviewed by AI.

\revision{Over a three-day period, we analyzed \participants{} submissions, evenly divided between the refactoring and Scrum contexts. The balanced set prevents one protocol from dominating the aggregate results, but the protocols were not experimental conditions and the study does not support causal comparisons between topics. Our focus is the workflow's operational execution, participant perceptions, and methodological limitations.}
\revision{Respondents generally rated the experience positively, while the item comparing AI and human interviewing produced a mixed distribution. The artifact audit showed that most submissions followed the expected structured-synthesis format, but also revealed incomplete, unusable, and protocol-inconsistent submissions. Reported limitations included generic questions, limited sensitivity, insufficiently deep probing, missed human interaction, privacy concerns, and repetitive questions.}
\revision{Accordingly, we position AI interviewers as a complementary instrument for short, self-administered interviews rather than as a replacement for human researchers, and we distinguish participant acceptability and structural artifact completeness from untested claims about richness, fidelity, and equivalence to human interviews.}
This article makes the following contributions:
\begin{itemize}
    \item \revision{We document an end-to-end workflow for self-administered, voice-based AI-conducted interviews in two ESE contexts and report participant perceptions, operational observations, and a structural audit of \participants{} submitted artifacts.}

    \item \revision{We characterize the submitted outputs as AI-mediated artifacts, distinguishing evidence about workflow execution and acceptability from untested claims about transcript fidelity, response richness, and equivalence to human-conducted interviews.}

    \item \revision{We derive practical lessons concerning pilot testing, participant instructions, natural-language handling, platform variability, privacy, artifact verification, and human oversight.}
\end{itemize}

\revision{The remainder of this article is organized as follows. Section~\ref{sec:study} describes the study design, including the research questions, study contexts, participants, ethical and privacy considerations, AI interviewer configuration, interview protocols, collected data, analysis procedure, and replication package. Section~\ref{sec:results} presents and discusses the findings regarding operational feasibility, participant perceptions, benefits, limitations, lessons learned, and the descriptive content of the submitted interview artifacts. Section~\ref{sec:threats} discusses the threats to validity, and Section~\ref{sec:related-work} positions our study in relation to prior research on software engineering interviews and AI-supported qualitative data collection. Finally, Section~\ref{sec:conclusion} summarizes the main findings, practical implications, limitations, and directions for future work.}

\section{Study Design}
\label{sec:study}
\begin{figure*}[t]
\centering
\includegraphics[width=1.0\textwidth]{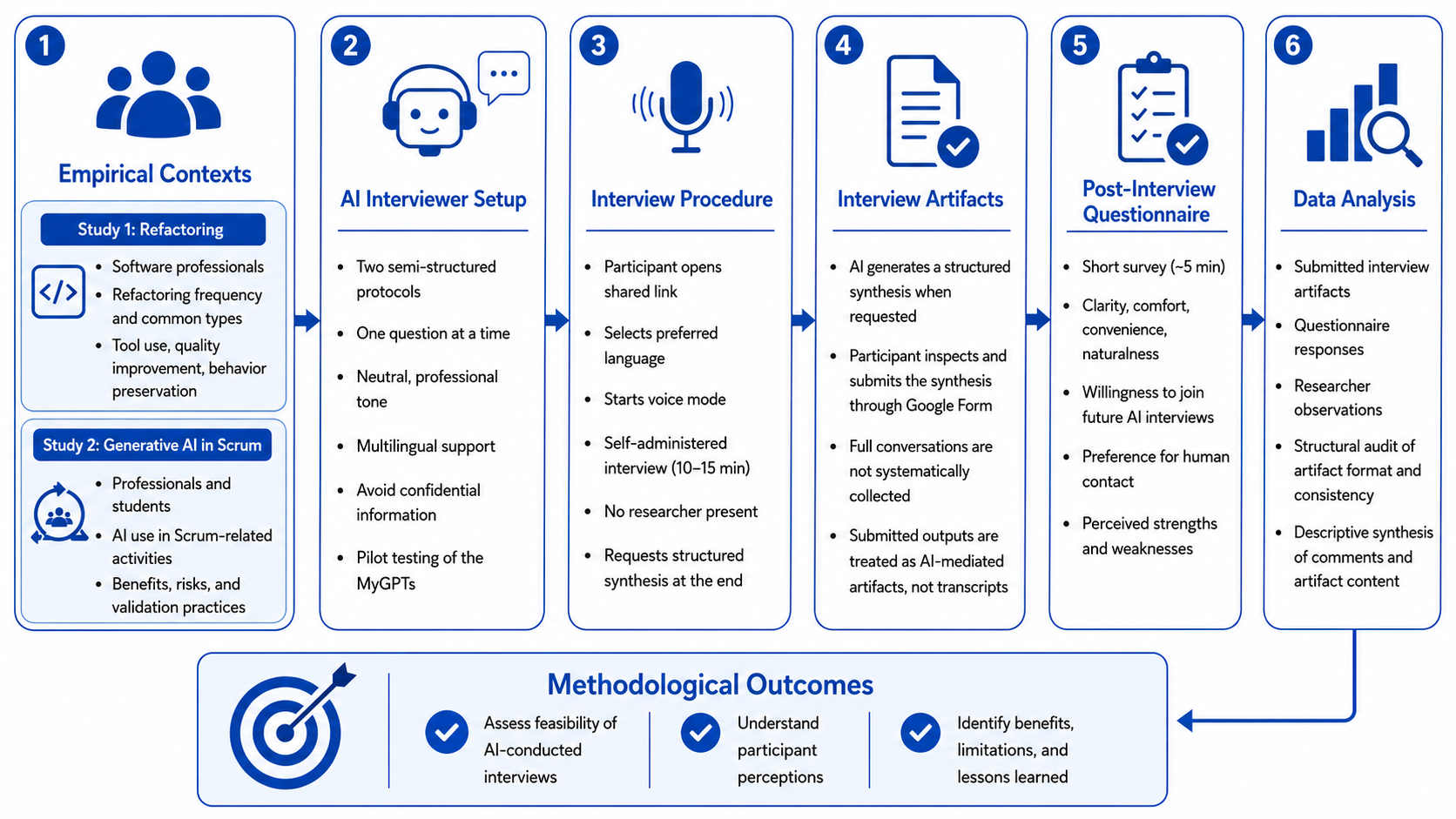}
\caption{Overview of the study design. Participants completed a self-administered voice-based interview with an AI interviewer, generated an interview artifact from the conversation, submitted the artifact and questionnaire responses through a post-interview form, and provided data for descriptive and qualitative analysis.}
\label{fig:technique}
\end{figure*}

\revision{This section describes an exploratory experience study of a self-administered AI-interview workflow in ESE.} We evaluated the approach in two interview studies: one on refactoring practices and another on the use of generative AI in Scrum-related activities. Figure~\ref{fig:technique} summarizes the overall workflow, from participant recruitment and AI interviewer setup to interview execution, artifact generation, questionnaire responses, and analysis.

\begin{figure*}[!t]
    \centering
    \includegraphics[width=1.0\textwidth]{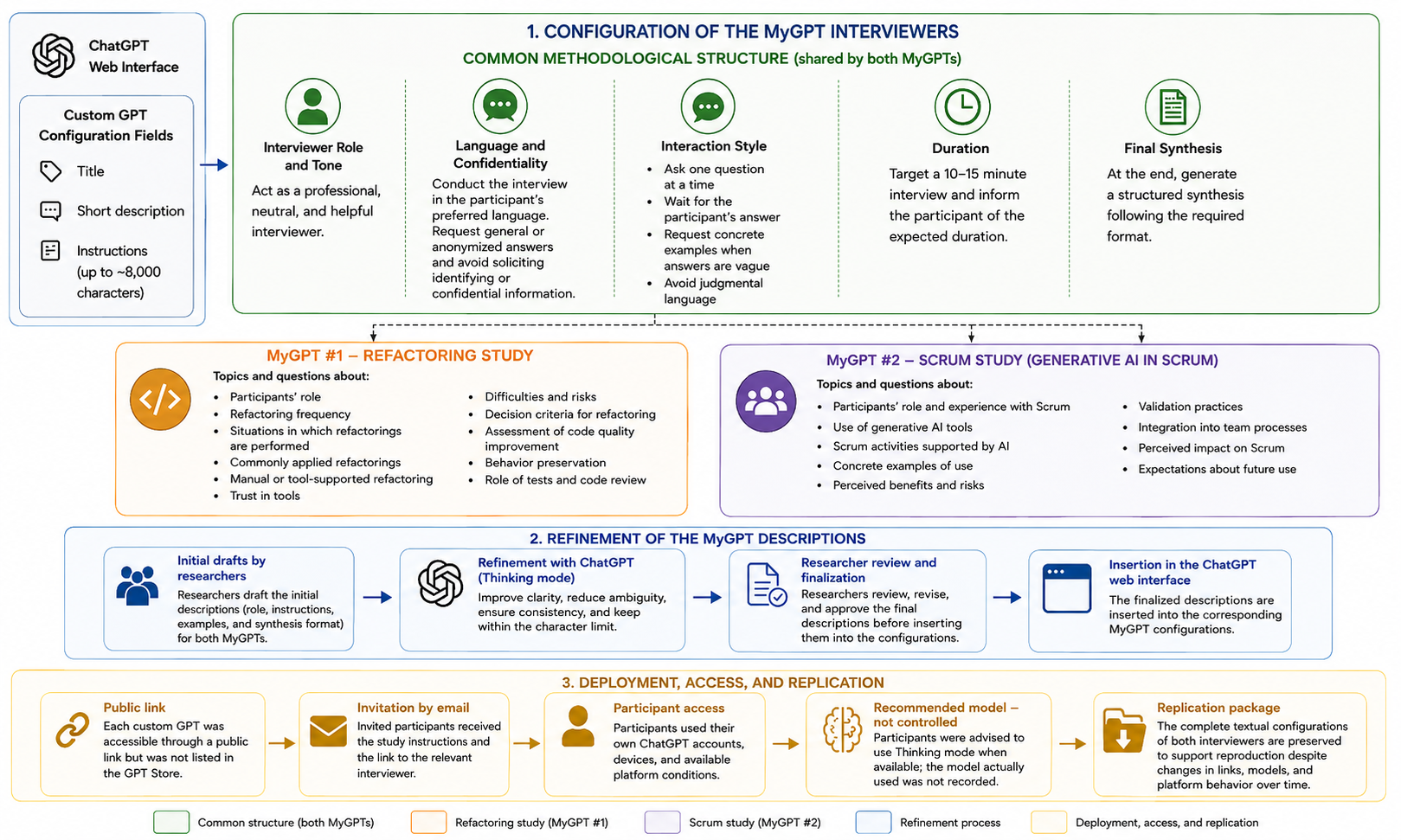}
    \caption{\revision{Workflow for configuring, refining, deploying, and supporting the replication of the MyGPT-based AI interviewers. The figure summarizes the shared methodological structure, study-specific components, AI-assisted prompt refinement followed by researcher review, participant access conditions, the non-controlled model setting, and the replication strategy.}}
    \label{fig:mygpt-workflow}
\end{figure*}

\revision{Figure~\ref{fig:mygpt-workflow} details the configuration,
refinement, deployment, participant-access, and replication workflow of the
MyGPT interviewers, complementing the high-level study overview in
Figure~\ref{fig:technique}.}

\subsection{Research Goal and Questions}
\label{subsec:rqs}

\revision{We state the goal of this study following the Goal-Question-Metric (GQM) template~\cite{Basili1994}: \emph{analyze} a MyGPT-based, self-administered interview workflow \emph{for the purpose of} characterizing it \emph{with respect to} its operational execution, participant perceptions, and methodological opportunities and limitations, \emph{from the viewpoint of} empirical software engineering researchers, \emph{in the context of} two ESE studies on refactoring practices and the use of generative AI in Scrum-related activities.} \revision{The study does not test whether AI-conducted interviews are equivalent or superior to human-conducted interviews.} We address the following research questions (RQs):

\begin{description}
\item[\textbf{RQ$_1$}] \revision{What evidence of operational feasibility and what practical difficulties emerged from the AI-conducted interview workflow?}

\item[\textbf{RQ$_2$}] How do participants perceive the experience of being interviewed by AI?

\item[\textbf{RQ$_3$}] What benefits, limitations, and lessons emerge from using AI as an interviewer in ESE?
\end{description}

\revision{RQ$_1$ concerns successful execution among the submissions and the practical problems observed in access, voice interaction, natural-language switching, artifact generation, and submission.} \revision{Because the analysis uses completed submissions and does not use invitation, start, or abandonment logs, RQ$_1$ does not estimate a population response or completion rate.} RQ$_2$ focuses on participants' perceptions, including comfort, clarity, convenience, naturalness, willingness to participate again, and preference for or against human contact. We address this question using descriptive statistics from the post-interview questionnaire and participants' open-ended responses. \revision{RQ$_3$ synthesizes methodological opportunities and limitations from questionnaire responses, submitted AI-mediated artifacts, open-ended comments, and procedural observations; it does not assess transcript fidelity or equivalence to human interviews.}

\subsection{Study Context}
\label{subsec:context}

We evaluated the AI interviewer in the context of two ESE studies. The first study investigated refactoring practices in real software projects. In this study, participants were asked about how often they perform refactorings, which refactorings they commonly apply, whether they use automated tools, what difficulties and risks they face, how they evaluate improvements in code quality, and how they verify that behavior has been preserved.
The second study investigated the use of generative AI in Scrum-related activities. In this study, participants were asked about how tools such as ChatGPT, Copilot, Gemini, Claude, DeepSeek, or similar systems are used in Scrum management activities such as backlog refinement, user story writing, acceptance criteria definition, sprint planning, documentation, communication, prioritization, decision support, testing, and quality-related tasks.

\revision{These two contexts were selected to provide breadth across a technical maintenance topic and a process-oriented topic involving generative AI.} \revision{This choice allows us to report whether the workflow could be deployed in two complementary ESE settings, but it reduces the depth with which any single domain or interview process can be evaluated.} \revision{The protocols are therefore treated as two application contexts rather than as controlled experimental conditions, and aggregate findings may reflect differences in topic, participant background, protocol design, or familiarity with AI.}
\revision{We prioritized breadth over depth in this first deployment because our primary goal was to examine whether the same self-administered workflow could be executed in more than one ESE context before investing in a deeper single-domain evaluation.} \revision{An in-depth, single-topic assessment of probing quality, follow-up behavior, clarification, elicitation of concrete examples, and adaptation to participant expertise requires full transcripts and, ideally, a human-interviewer baseline, neither of which was part of this study; we identify this as future work in Section~\ref{sec:conclusion}.}

Participants completed a short self-administered interview, lasting approximately 10--15 minutes, through a custom AI interviewer available via a shareable link. The interviewer was designed to use voice interaction, ask one question at a time, and request concrete examples when appropriate. \revision{Table~\ref{tab:protocol-summary} summarizes the two interview protocols, the collected artifacts, the shared questionnaire constructs, and their mapping to the RQs.}

\begin{table*}[t]
\centering
\small
\caption{\revision{Summary of the two interview protocols, collected artifacts, questionnaire constructs, and RQ mapping.}}
\label{tab:protocol-summary}
\begin{tabular}{@{}p{0.18\textwidth}p{0.37\textwidth}p{0.37\textwidth}@{}}
\toprule
\revision{\textbf{Element}} & \revision{\textbf{Refactoring protocol}} & \revision{\textbf{Scrum and generative AI protocol}} \\
\midrule
\revision{Study context} & \revision{Technical maintenance practices in real software projects.} & \revision{Process-oriented use of generative AI in Scrum-related activities.} \\
\revision{Core topics} & \revision{Roles, refactoring frequency and types, tool use, trust, risks, code quality, behavior preservation, tests, and code review.} & \revision{Roles, AI tools, supported Scrum activities, examples, benefits, risks, validation, process integration, impact, and future use.} \\
\revision{Submitted artifact} & \revision{AI-generated synthesis organized around participant profile, refactoring practices, tool support, risks, validation, and possible codes.} & \revision{AI-generated synthesis organized around participant profile, AI use, Scrum activities, benefits, risks, validation, governance, impact, and possible codes.} \\
\revision{Analyzed respondents} & \revision{\participantsRefactoring{} submissions.} & \revision{\participantsScrum{} submissions.} \\
\revision{Shared questionnaire} & \multicolumn{2}{p{0.74\textwidth}}{\revision{Clarity, comfort, naturalness, ability to express experiences, pace, comparative orientation toward AI or human interviewing, effect of human absence, willingness to participate again, perceived strengths, weaknesses, and open comments.}} \\
\revision{RQ mapping} & \multicolumn{2}{p{0.74\textwidth}}{\revision{RQ$_1$: operational execution and difficulties; RQ$_2$: participant perceptions; RQ$_3$: methodological opportunities, limitations, and lessons.}} \\
\bottomrule
\end{tabular}
\end{table*}

\subsection{Participants}
\label{subsec:participants}

\revision{Participants were invited by email and received the link for the interview protocol relevant to the study in which they were participating.} \revision{Recruitment followed a convenience sampling strategy: invitations were sent by email to individuals expected to have experience relevant to one of the two topics, and each invitee received only the link for the corresponding study.} \revision{Assignment to a protocol was therefore not randomized; it was determined by the study for which the participant was recruited, based on self-reported experience with that topic, and the two groups were recruited independently.} \revision{Because we did not systematically track the total number of invitations sent, we cannot report a response rate, and the sample may not be representative of the broader population of software practitioners and students.} \revision{The inclusion criteria were consent to the use of anonymized data, submission of the interview artifact and questionnaire, and self-reported experience relevant to the interview topic; submissions that did not meet these criteria were not analyzed.} \revision{We analyzed \participants{} questionnaire responses from individuals who self-reported experience relevant to at least one interview topic: \participantsRefactoring{} completed the refactoring protocol and \participantsScrum{} completed the Scrum and generative AI protocol.} \revision{The study was not designed as a randomized comparison between these protocols; the balanced split is used only to avoid one context dominating the aggregate perception results.} \revision{The refactoring group included professionals with experience in development, maintenance, testing, code review, architecture, or technical leadership, whereas the Scrum group included professionals and students with experience in Scrum-related software development activities.}

Participation was voluntary. Before starting the interview, participants received instructions explaining the purpose of the study, the expected duration, the use of an AI interviewer, and the need to complete a short post-interview form. Because the interviews involved professional experiences, participants were also instructed not to disclose confidential or identifying information. \revision{The analysis is based on valid completed submissions and does not use a denominator covering invitations, interview starts, or incomplete attempts; consequently, we do not report a response or end-to-end completion rate.}

All participants who submitted valid questionnaire responses self-reported living in \country{}. \revision{The experience distribution (Figure~\ref{fig:additional-perceptions}e) included 19 participants with 10 or more years (28.8\%), 16 with 2--3 years (24.2\%), 13 with 4--5 years (19.7\%), 10 with 6--7 years (15.2\%), six with 0--1 year (9.1\%), and two with 8--9 years (3.0\%).} \revision{Using broader experience bands, the refactoring group comprised four participants with 0--1 year, nine with 2--3 years, seven with 4--5 years, four with 6--9 years, and nine with 10 or more years; the corresponding counts in the Scrum group were two, seven, six, eight, and ten.} \revision{Professional-role details were available only in some AI-mediated artifacts rather than as standardized questionnaire variables, so we report broad eligibility categories and do not perform role-based subgroup analyses.} \revision{The present analysis does not stratify participants by industry sector, prior experience as research-interview participants, prior use of voice mode, or detailed familiarity with AI interviewers; these characteristics may influence comfort and perceived usability.}

\subsection{Ethical and Privacy Considerations}
\label{subsec:ethical}

\revision{Because the study involved human participants and an external AI platform, we adopted safeguards focused on informed participation, voluntary submission, confidentiality, and anonymization.} \revision{Participants were informed that the interview would be conducted by an AI-based interviewer and that the research team would receive only the questionnaire responses and interview artifact that they voluntarily submitted through the post-interview Google Form.} \revision{Nevertheless, participants interacted with ChatGPT through their own accounts, and the research team did not control platform-side processing, storage, account settings, or retention; avoiding confidential disclosure was therefore an important but incomplete safeguard.}
\revision{Concretely, everything a participant said or typed during the interview was processed by the external platform and may have been retained in the participant's conversation history or used according to the platform's data policies and the participant's individual account settings (for example, settings related to using conversations for model training).} \revision{Although participants knew that the interaction occurred on an external AI platform through their own accounts, the study instructions did not include a detailed description of these platform-side retention and training policies.} \revision{Future deployments should include an explicit platform-privacy notice describing what data the platform may process or retain and, when applicable, instructions for reviewing account-level privacy settings before starting the interview.}

The participant-facing instructions emphasized that participants should not provide confidential or identifying information about companies, clients, projects, systems, products, or individuals. The AI interviewer was also instructed to avoid collecting such information and, when necessary, to ask participants to answer in a general or anonymized way. Before analysis and reporting, the collected data were inspected to remove or generalize potentially identifying information.


\subsection{AI Interviewer Setup}
\label{subsec:ai}

The AI interviewer was implemented using the GPT feature available through the ChatGPT web interface. GPTs allow users to create customized conversational agents, commonly referred to as MyGPTs~\cite{openai2026creatinggpts}, by defining a title, a short description, and a set of instructions that specify the agent's expected behavior. At the time of our study, the instruction field supported up to approximately 8,000 characters. This constraint influenced how we designed the interviewer prompts: the instructions had to be sufficiently detailed to guide the interview, but concise enough to fit within the available character limit.

We created two separate MyGPTs, one for each interview study. The first MyGPT was configured for interviews about refactoring practices, and the second was configured for interviews about the use of generative AI in Scrum-related activities. Both MyGPTs shared a common methodological structure. This common part specified the interviewer's role, expected tone, language behavior, confidentiality safeguards, interaction style, duration, and final synthesis format. In particular, the AI interviewer was instructed to ask one question at a time, wait for the participant's answer, avoid judgmental language, request concrete examples when answers were vague, avoid collecting confidential or identifying information, and conduct the interview in the participant's preferred language.

Each MyGPT also included a study-specific component. The refactoring interviewer covered participants' roles, refactoring frequency, commonly applied refactorings, manual or tool-supported refactoring, trust in tools, difficulties and risks, quality assessment, behavior preservation, tests, and code review. The Scrum interviewer covered participants' roles, generative AI tools, Scrum activities supported by AI, concrete examples, perceived benefits and risks, validation practices, process integration, perceived impact on Scrum, and expectations about future use.
The initial descriptions of the two MyGPTs were drafted by the researchers and then refined with support from ChatGPT in Thinking 5.5 mode. 
\revision{We used this AI-assisted refinement step to improve clarity, reduce
ambiguity, promote consistency between the two configurations, and keep the
instructions within the available character limit. The researchers subsequently
reviewed, revised, and approved the resulting descriptions before inserting them
into the corresponding MyGPT configurations through the ChatGPT web interface.}

\revision{Before deployment, the researchers conducted a technical and protocol-oriented pilot of the MyGPTs to check whether the interviewer followed the intended behavior.} \revision{This pilot used the OpenAI Plus account employed to create the MyGPTs and was not analyzed as a separate participant study or as evidence of interview-data quality.} \revision{The pilot covered the participant-facing procedure from the participant's perspective---access through the shared link (including with free accounts), voice interaction, natural-language switching, protocol adherence, and generation of the final synthesis---but it was researcher-led: no external pilot participants completed the full end-to-end package, including the recruitment instructions and the post-interview questionnaire, before deployment.} \revision{We recommend that future deployments include a small external pilot of the complete workflow.} During this pilot, we verified whether the interviewer started by asking the preferred language, explained the purpose of the interview, asked one question at a time, avoided collecting confidential information, used short follow-up questions when appropriate, respected the expected duration, and generated a structured synthesis aligned with the study goals. We also tested the interaction in three different languages to assess whether the interviewer could switch languages consistently and produce the final synthesis in the language used during the interview. We also checked whether participants could access and use the shared MyGPT links with free ChatGPT accounts, which worked in our study context. This pilot step was important because participants do not see the internal MyGPT instructions; they only interact with the resulting conversational behavior. Researchers need to verify in advance whether the configured AI interviewer behaves as intended from the participant's perspective and under the access conditions expected for participants.

For deployment, each MyGPT was made accessible through a public link, but it was not published in the GPT Store. The links were sent by email to invited participants together with the study instructions. Participants accessed the interviewers through their own ChatGPT accounts, subject to the access conditions available on the platform at the time of the study. The recommended model for the interaction was the latest available model at the time of the study, ChatGPT in Thinking 5.5 mode, when available in the participant's interface. The shared GPT pages displayed the account name associated with their creation. The model was recommended, but the model actually used by each participant was not recorded or controlled.
The replication package~\cite{artefatos} includes the complete textual descriptions used to configure each MyGPT.

\subsection{AI Interviewer Protocols}
\label{subsec:protocols}

The two AI interviewer protocols were designed as semi-structured interview guides. They were not intended to force a rigid questionnaire-like interaction. Instead, they specified a sequence of core questions and allowed the AI interviewer to ask short follow-up questions when needed. The protocols also instructed the AI to prioritize concrete experiences, ask for examples, and avoid introducing assumptions or leading participants toward particular answers.
\revision{The first interaction in both protocols asked participants to indicate their preferred natural language; throughout this article, \emph{language} in the interview context refers to a human language rather than a programming language.} \revision{After the participant selected or requested a natural language, the AI interviewer explained the purpose of the interview and continued in that language.} \revision{The protocols also instructed the AI to generate the final synthesis in the natural language used during the interview.} \revision{This design offered multilingual interaction, but the study was not a cross-language validation and did not evaluate semantic equivalence, translation accuracy, or summary fidelity across languages.}

Both protocols included an explicit closing step. Near the end of the interview, the AI was instructed to inform the participant that it would ask the final question. After the participant answered, the AI closed the interview and generated a structured synthesis based only on what had been said. The synthesis template differed between the two studies. The refactoring synthesis included blocks such as participant profile, frequency of refactoring, types of refactorings mentioned, use of tools, confidence in tools, difficulties and risks, quality assessment, behavior preservation, and possible qualitative codes. The Scrum synthesis included blocks such as participant profile, AI tools mentioned, frequency of use, Scrum activities supported by AI, concrete example cited, perceived benefits, risks and limitations, validation practices, governance or confidentiality concerns, perceived impact on Scrum, future opportunities, and possible qualitative codes.

\subsection{Post-Interview Questionnaire}
\label{subsec:questionario}

After completing the interview, participants were asked to answer a short questionnaire about their experience of being interviewed by AI. The questionnaire was administered through Google Forms and was designed to be completed in approximately five minutes. It included a consent question followed by closed-ended questions about clarity, comfort, convenience, naturalness, willingness to participate in future AI-conducted interviews, and confidence in sharing opinions with an AI interviewer. It also included questions about perceived strengths and weaknesses of the AI-conducted interview process.

The questionnaire captured participants' perceptions of the interview method, rather than the substantive content of their answers about refactoring or Scrum. \revision{It therefore supports claims about perceived clarity, comfort, convenience, and acceptability, but not claims about the depth, completeness, accuracy, or analytical value of the qualitative evidence.} \revision{The questionnaire also included a self-reported comparative item but no human-interview control group.} \revision{Because the wording of that item referred to how participants felt they answered while its scale endpoints referred to preference for AI or a human interviewer, we treat it as a broad comparative orientation and not as a precise measure of either response quality or interviewer preference.}

\subsection{Collected Data}
\label{subsec:data}

The study produced three sources of data: interview artifacts voluntarily submitted through the post-interview form, questionnaire responses submitted through the same form, and researchers' observations about the execution of the procedure, including issues related to language switching, voice mode, access to the AI interviewer, artifact generation, and submission.

\revision{In total, we analyzed \participants{} questionnaire responses from participants who consented to the use of their anonymized data: \participantsRefactoring{} from the refactoring interview and \participantsScrum{} from the Scrum and generative AI interview.} \revision{We use these data as evidence about successful workflow execution among completers, participant perceptions, and observed methodological limitations.}
\revision{The research team did not systematically collect the participants' full conversations with the MyGPT; the standardized dataset consists of materials intentionally submitted through the Google Form and procedural observations.} \revision{Participants could inspect the generated synthesis before submitting it; however, we do not treat submission as a formal member check of completeness or fidelity.} \revision{Accordingly, the submitted artifacts are neither verbatim transcripts nor participant-validated representations of the full conversation.}

\subsection{Analysis Procedure}
\label{subsec:analysis}

We analyzed the data using a combination of descriptive and qualitative procedures. Questionnaire responses were summarized using descriptive statistics to characterize participants' perceptions of the AI-conducted interview experience. For closed-ended items, we analyzed response distributions to identify general tendencies regarding comfort, clarity, convenience, naturalness, acceptance, perceived strangeness, and preference for human contact.

\revision{The interview artifacts and open-ended questionnaire responses were inspected descriptively to identify recurring benefits, limitations, procedural issues, and topic-relevant content. We conducted a post-hoc structural audit of all \participants{} submitted artifacts using explicit descriptive criteria. An artifact was classified as an expected structured synthesis when it contained the study-specific participant-profile section and multiple protocol sections; otherwise, it was classified as a conversation-style excerpt, topic-relevant but unstructured answers, or unusable content. We separately recorded whether an explicit or functionally equivalent qualitative-code block was present, inferred the artifact's predominant natural language, counted whitespace-delimited words, and checked whether its topic was consistent with the protocol reported in the questionnaire. These checks assess submission format and topical consistency rather than richness, completeness, or fidelity to the original conversation.} \revision{One researcher conducted the structural audit, and the research team discussed the ambiguous and protocol-inconsistent cases. Because the audit was not independently replicated by a second coder, its classifications may be subject to researcher judgment.} \revision{Any ``possible qualitative codes'' generated by the MyGPT were treated as part of the AI-mediated artifact and not as independently validated researcher codes.} \revision{The quantified outcomes of this audit are reported in Section~\ref{subsec:rq1-feasibility}.}

We then inspected the data to identify recurring benefits, limitations, and lessons learned from the use of AI as an interviewer. The categories were organized around the research questions, distinguishing evidence related to feasibility, participant experience, and methodological implications. Examples of categories include scheduling flexibility, multilingual support, standardization of the interview flow, convenience, comfort, perceived lack of human contact, technical difficulties, dependence on participant instructions, loss of detail in AI-generated summaries, and limitations for artifact-rich discussions.

\revision{The analysis focused on the methodological experience of using AI to conduct interviews rather than on producing a complete theory of refactoring practices or generative AI use in Scrum.} \revision{Because no full-transcript comparison, systematic member checking, human-interviewer control condition, or formal richness assessment was performed, the artifact inspection cannot establish fidelity, completeness, or equivalence to human-conducted interviews.} \revision{The results should therefore be interpreted as an exploratory experience report.}

\revision{The replication package~\cite{artefatos} includes the post-interview questionnaire, anonymized questionnaire responses in CSV format, the textual configurations of the two MyGPT interviewers, workflow documentation, and the data used to reproduce the descriptive questionnaire and artifact-audit results.} \revision{Full MyGPT conversations were not systematically collected and are therefore not part of the standardized dataset.} \revision{All released participant data were anonymized by removing or generalizing names of people, companies, clients, projects, products, systems, and other potentially identifying information.} \revision{Because AI platforms may change over time, we treat the prompts, questionnaire, workflow documentation, and data-collection procedure as the primary replicable artifacts.}

\section{Results and Discussion}
\label{sec:results}

\revision{This section reports operational observations and participant perceptions from \participants{} responses, evenly distributed across the refactoring and Scrum protocols.} \revision{The results are conditional on respondents who completed and submitted the workflow and should not be interpreted as a response-rate estimate or as a direct comparison with human-conducted interviews.} \revision{Participants generally rated comfort, clarity, pace, and willingness to participate again positively, while the comparative item and reported limitations indicate more mixed views about human absence, probing depth, privacy, and interaction quality.} The following subsections discuss these results in relation to our RQs.

\begin{figure*}[t]
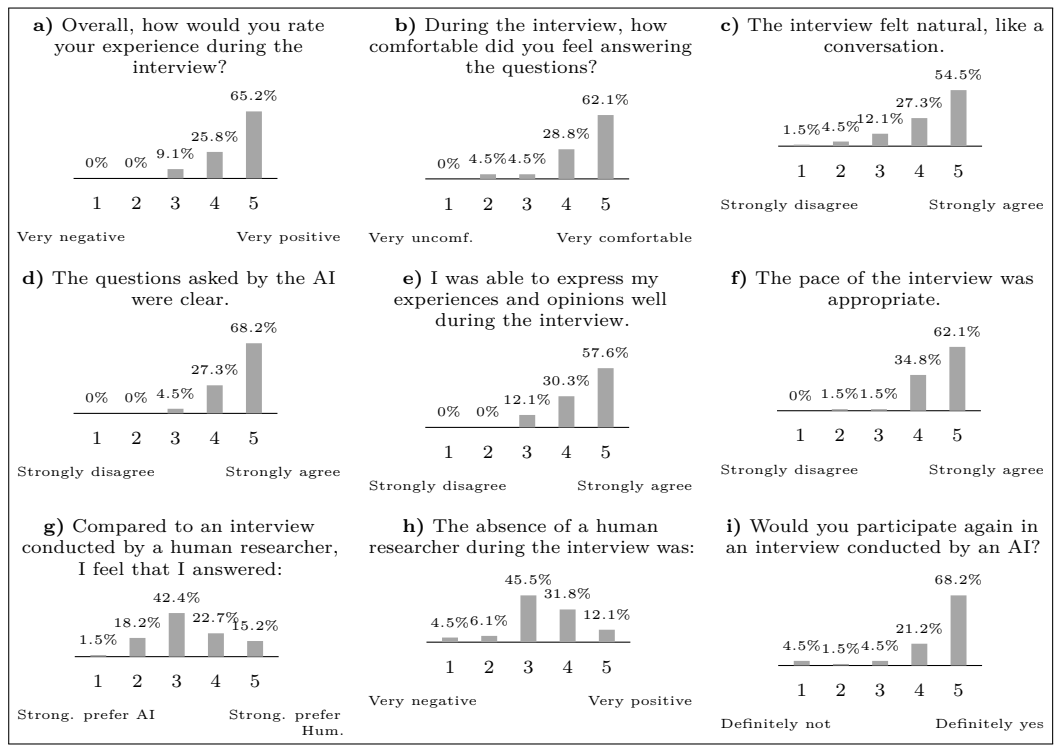

\centering
\fbox{%
\begin{minipage}{0.97\textwidth}
\centering

\likertbarcompactlr{a)}{Overall, how would you rate your experience during the interview?}
{Very negative}
{Very positive}
{(1,0) (2,0) (3,9.1) (4,25.8) (5,65.2)}
\hfill
\likertbarcompactlr{b)}{During the interview, how comfortable did you feel answering the questions?}
{Very uncomf.}
{Very comfortable}
{(1,0) (2,4.5) (3,4.5) (4,28.8) (5,62.1)}
\hfill
\likertbarcompactlr{c)}{The interview felt natural, like a conversation.}
{Strongly disagree}
{Strongly agree}
{(1,1.5) (2,4.5) (3,12.1) (4,27.3) (5,54.5)}

\vspace{3mm}

\likertbarcompactlr{d)}{The questions asked by the AI were clear.}
{Strongly disagree}
{Strongly agree}
{(1,0) (2,0) (3,4.5) (4,27.3) (5,68.2)}
\hfill
\likertbarcompactlr{e)}{I was able to express my experiences and opinions well during the interview.}
{Strongly disagree}
{Strongly agree}
{(1,0) (2,0) (3,12.1) (4,30.3) (5,57.6)}
\hfill
\likertbarcompactlr{f)}{The pace of the interview was appropriate.}
{Strongly disagree}
{Strongly agree}
{(1,0) (2,1.5) (3,1.5) (4,34.8) (5,62.1)}

\vspace{3mm}

\likertbarcompactlr{g)}{Compared to an interview conducted by a human researcher, I feel that I answered:}
{Strong. prefer AI}
{Strong. prefer Hum.}
{(1,1.5) (2,18.2) (3,42.4) (4,22.7) (5,15.2)}
\hfill
\likertbarcompactlr{h)}{The absence of a human researcher during the interview was:}
{Very negative}
{Very positive}
{(1,4.5) (2,6.1) (3,45.5) (4,31.8) (5,12.1)}
\hfill
\likertbarcompactlr{i)}{Would you participate again in an interview conducted by an AI?}
{Definitely not}
{Definitely yes}
{(1,4.5) (2,1.5) (3,4.5) (4,21.2) (5,68.2)}

\end{minipage}%
}
\caption{\revision{Results of the post-interview questionnaire with \participants{} responses. Each chart shows the percentage distribution across the five-point Likert scale. In panel (g), the question wording referred to how participants felt they answered, whereas the endpoints referred to preference for AI or a human interviewer; we therefore interpret this item only as a broad comparative orientation.}}
\label{fig:post-interview-questionnaire}
\end{figure*}

\subsection{\revision{RQ$_1$: Operational Feasibility and Practical Difficulties}}
\label{subsec:rq1-feasibility}

\revision{Our first RQ examines evidence of operational feasibility and practical difficulties in the self-administered workflow.} \revision{All \participants{} analyzed submissions included a non-empty artifact field, and the post-hoc audit showed that 92.4\% (61/66) contained a structured synthesis with the expected sections; 72.7\% (48/66) also included an explicit or functionally equivalent qualitative-code block.} \revision{Of the remaining five submissions, three (4.5\%, 3/66) contained conversation-style excerpts of the interview without the final structured synthesis, one (1.5\%, 1/66) contained topic-relevant answers pasted without the expected structure, and one (1.5\%, 1/66) did not contain usable interview content, consisting only of the assistant's generic offer to structure content.} \revision{The submitted artifacts had a median length of 442 words (interquartile range 258--713; range 66--2,371), consistent with short, focused interviews.} \revision{As a separate consistency check, two artifacts (3.0\%, 2/66) contained clear refactoring content although the corresponding questionnaire records identified the Scrum protocol. Because the available data do not reveal whether the protocol response or the pasted artifact was incorrect, we did not reassign these records. We retained their questionnaire responses in the aggregate perception analysis, but excluded the two artifacts from protocol-specific content descriptions.} \revision{These deviations show that self-administered artifact submission can fail silently---participants may paste the wrong portion of the conversation, omit the expected synthesis, or submit content inconsistent with the selected protocol---and reinforce the need to inspect every submitted artifact before analysis.} \revision{This evidence is conditional on successful completers: because the analysis does not use a complete log of invitations, starts, and abandonments, it cannot establish an overall completion rate.} \revision{For the analyzed respondents, the asynchronous design removed the need to coordinate a synchronous researcher--participant session and allowed completion at a self-selected time.}

\revision{Perceived pace and clarity also support the usability of the procedure among completers: 97.0\% (64/66) rated the pace positively and 95.5\% (63/66) rated the questions as clear.} \revision{These ratings describe participant perceptions rather than objective measures of pacing or protocol adherence.} \revision{Voice interaction enabled oral participation and subsequent artifact generation, but it also introduced technical risks.} \revision{For example, one participant reported that a cough was interpreted as an answer, illustrating how audio input and speech recognition can alter the interview flow.}

\revision{Natural-language handling was both an opportunity and a source of variability.} \revision{Participants could request a preferred natural language, but some needed to repeat a language-change request and the AI did not always respond consistently.} \revision{Participants also used their own accounts and available models, so differences in account tier, interface, or model access may have influenced the interaction and the exact model could not be standardized across interviews.} \revision{The natural language used in each interview was not recorded as a structured variable; however, 65/66 submitted artifacts were written predominantly in Portuguese and one in English. Because the protocol instructed the AI to generate the synthesis in the interaction language, artifact language provides a useful but imperfect proxy.} \revision{The multilingual capability was piloted in three languages, but the submitted artifacts provide little empirical evidence about multilingual use in practice; we also did not evaluate translation quality or semantic equivalence across natural languages.} \revision{Future studies should record the interview language per participant and recruit linguistically diverse samples to enable language-related analyses.}

\revision{Artifact generation and submission were operationally successful for the valid submissions, and the structured output was immediately available after the interaction.} \revision{However, we did not measure net researcher time savings, and the output may summarize, reorganize, or omit details from the original conversation.} \revision{Because the study did not systematically collect full transcripts or require member checking, artifact availability should not be interpreted as evidence of fidelity or qualitative-data equivalence.}

\begin{tcolorbox}[
    colback=gray!8,
    colframe=black!60,
    title={\faLightbulb\quad Finding for RQ$_{1}$},
    fonttitle=\bfseries,
    sharp corners,
    boxrule=0.6pt
]
\noindent
\revision{Among the \participants{} submissions, respondents completed the self-administered workflow, and 92.4\% (61/66) of the submitted artifacts followed the expected structured-synthesis format; the remaining submissions contained conversation excerpts without the synthesis, unstructured answers, or unusable content. Two artifacts also conflicted with the protocol reported in the corresponding questionnaire record. This demonstrates operational viability among completers---not an overall completion rate or the methodological adequacy of the resulting data---and shows that submissions and protocol--artifact linkage must be individually verified. Clear instructions, precise form wording, reliable voice interaction, natural-language handling, and validation of generated artifacts remain necessary.}
\end{tcolorbox}

\subsection{RQ$_2$: Participants' Perceptions of Being Interviewed by AI}
\label{subsec:rq2-perceptions}

Our second RQ investigates how participants perceived the experience of being interviewed by AI. \revision{Figure~\ref{fig:post-interview-questionnaire} summarizes the distributions of the closed-ended questionnaire items.} \revision{Overall experience was rated positively by 90.9\% (60/66), 90.9\% (60/66) reported feeling comfortable, 97.0\% (64/66) rated the pace positively, and 89.4\% (59/66) indicated that they would participate again.} \revision{These results support acceptability among the analyzed respondents but do not indicate whether non-completers would report similar perceptions.}

\begin{table}[t]
\centering
\small
\caption{\revision{Positive questionnaire ratings (scores 4--5) by self-reported interview protocol. The groups were recruited independently and were not randomized experimental conditions; the table is descriptive.}}
\label{tab:perceptions-by-protocol}

\renewcommand{\arraystretch}{1.15}

\begin{tabular}{
    @{}
    L{0.46\textwidth}
    R{0.20\textwidth}
    R{0.20\textwidth}
    @{}
}
\toprule
\revision{\textbf{Questionnaire item}}
&
\multicolumn{1}{c}{\revision{\textbf{Refactoring ($n=33$)}}}
&
\multicolumn{1}{c}{\revision{\textbf{Scrum ($n=33$)}}}
\\
\midrule

\revision{Positive overall experience}
& \revision{29/33 (87.9\%)}
& \revision{31/33 (93.9\%)}
\\

\revision{Comfortable answering}
& \revision{30/33 (90.9\%)}
& \revision{30/33 (90.9\%)}
\\

\revision{Interview felt natural}
& \revision{25/33 (75.8\%)}
& \revision{29/33 (87.9\%)}
\\

\revision{Questions were clear}
& \revision{30/33 (90.9\%)}
& \revision{33/33 (100.0\%)}
\\

\revision{Able to express experiences and opinions}
& \revision{27/33 (81.8\%)}
& \revision{31/33 (93.9\%)}
\\

\revision{Pace was appropriate}
& \revision{31/33 (93.9\%)}
& \revision{33/33 (100.0\%)}
\\

\revision{Would participate again}
& \revision{31/33 (93.9\%)}
& \revision{28/33 (84.8\%)}
\\

\bottomrule
\end{tabular}
\end{table}

\begin{figure*}[!ht]
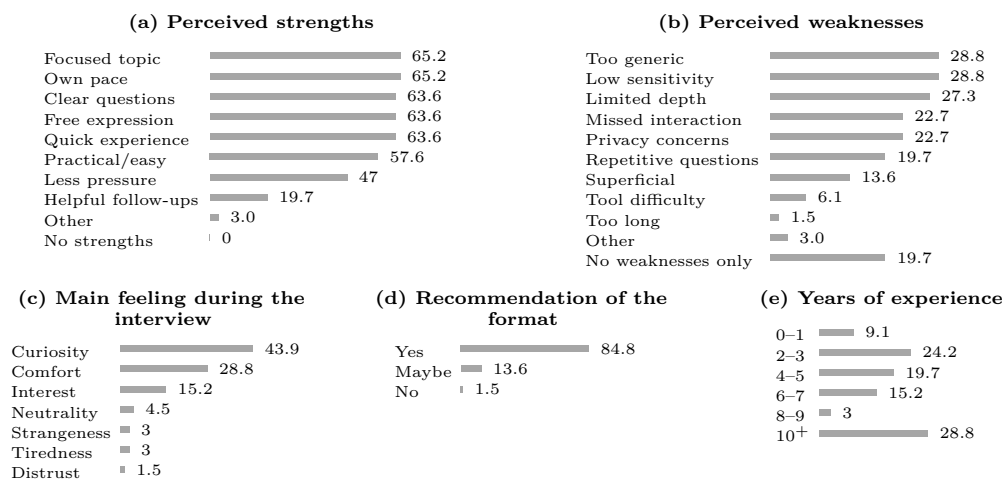

\centering

\begin{minipage}[t]{0.49\textwidth}
\centering
\scriptsize
\textbf{(a) Perceived strengths}

\vspace{1mm}

{\fontsize{6.5}{7}\selectfont
\begin{tabular}{@{}l@{\hspace{1.5mm}}l@{}}
\hbarrowpct{Focused topic}{65.2}{70}
\hbarrowpct{Own pace}{65.2}{70}
\hbarrowpct{Clear questions}{63.6}{70}
\hbarrowpct{Free expression}{63.6}{70}
\hbarrowpct{Quick experience}{63.6}{70}
\hbarrowpct{Practical/easy}{57.6}{70}
\hbarrowpct{Less pressure}{47}{70}
\hbarrowpct{Helpful follow-ups}{19.7}{70}
\hbarrowpct{Other}{3.0}{70}
\hbarrowpct{No strengths}{0}{70}
\end{tabular}
}
\end{minipage}
\hfill
\begin{minipage}[t]{0.49\textwidth}
\centering
\scriptsize
\textbf{(b) Perceived weaknesses}

\vspace{1mm}

{\fontsize{6.5}{7}\selectfont
\begin{tabular}{@{}l@{\hspace{1.5mm}}l@{}}
\hbarrowpct{Too generic}{28.8}{35}
\hbarrowpct{Low sensitivity}{28.8}{35}
\hbarrowpct{Limited depth}{27.3}{35}
\hbarrowpct{Missed interaction}{22.7}{35}
\hbarrowpct{Privacy concerns}{22.7}{35}
\hbarrowpct{Repetitive questions}{19.7}{35}
\hbarrowpct{Superficial}{13.6}{35}
\hbarrowpct{Tool difficulty}{6.1}{35}
\hbarrowpct{Too long}{1.5}{35}
\hbarrowpct{Other}{3.0}{35}
\hbarrowpct{No weaknesses only}{19.7}{35}
\end{tabular}
}
\end{minipage}

\vspace{3mm}

\begin{minipage}[t]{0.32\textwidth}
\centering
\scriptsize
\textbf{(c) Main feeling during the interview}

\vspace{1mm}

{\fontsize{6.5}{7}\selectfont
\begin{tabular}{@{}l@{\hspace{1.2mm}}l@{}}
\hbarrowpctsmall{Curiosity}{43.9}{50}
\hbarrowpctsmall{Comfort}{28.8}{50}
\hbarrowpctsmall{Interest}{15.2}{50}
\hbarrowpctsmall{Neutrality}{4.5}{50}
\hbarrowpctsmall{Strangeness}{3}{50}
\hbarrowpctsmall{Tiredness}{3}{50}
\hbarrowpctsmall{Distrust}{1.5}{50}
\end{tabular}
}
\end{minipage}
\hfill
\begin{minipage}[t]{0.32\textwidth}
\centering
\scriptsize
\textbf{(d) Recommendation of the format}

\vspace{1mm}

{\fontsize{6.5}{7}\selectfont
\begin{tabular}{@{}l@{\hspace{1.2mm}}l@{}}
\hbarrowpctsmall{Yes}{84.8}{100}
\hbarrowpctsmall{Maybe}{13.6}{100}
\hbarrowpctsmall{No}{1.5}{100}
\end{tabular}
}
\end{minipage}
\hfill
\begin{minipage}[t]{0.32\textwidth}
\centering
\scriptsize
\textbf{(e) Years of experience}

\vspace{1mm}

{\fontsize{6.5}{7}\selectfont
\begin{tabular}{@{}l@{\hspace{1.2mm}}l@{}}
\hbarrowpctsmall{0--1}{9.1}{40}
\hbarrowpctsmall{2--3}{24.2}{40}
\hbarrowpctsmall{4--5}{19.7}{40}
\hbarrowpctsmall{6--7}{15.2}{40}
\hbarrowpctsmall{8--9}{3}{40}
\hbarrowpctsmall{10$^+$}{28.8}{40}
\end{tabular}
}
\end{minipage}

\caption{\revision{Additional perceptions about the AI-led interview and respondents' experience profile. Panels (a) and (b) summarize multiple-selection questions; therefore, percentages do not sum to 100\% because participants could select more than one option. Panels (c), (d), and (e) summarize single-choice questions.}}
\label{fig:additional-perceptions}
\end{figure*}

\revision{Table~\ref{tab:perceptions-by-protocol} presents the protocol-stratified distributions, which show that the positive aggregate results were present in both groups, although they were not identical. The Scrum group had descriptively higher positive ratings for naturalness, clarity, ability to express experiences, and pace, whereas the refactoring group had a higher proportion willing to participate again. Because protocol assignment was not randomized and the groups differed in topic and participant composition, these differences should not be attributed to the protocol or interview topic.} \revision{For the ambiguous comparative item, both groups had 14/33 midpoint responses; 13/33 refactoring and 12/33 Scrum respondents leaned toward the human-interviewer endpoint, while 6/33 and 7/33, respectively, leaned toward the AI-interviewer endpoint.}

\revision{Question clarity received positive ratings from 95.5\% (63/66).} \revision{In the multi-selection strengths question, 63.6\% (42/66) selected ``\textit{The questions were clear},'' and 65.2\% (43/66) selected ``\textit{The AI kept the interview focused on the topic}.''}

\revision{Naturalness received positive ratings from 81.8\% (54/66), a lower proportion than clarity or pace.} \revision{One open-ended comment illustrates that the artificial nature of the interaction remained salient for some participants: a participant reported that it felt strange to talk to a robot and that the voice tone was not natural.}

\revision{The ability to express experiences and opinions received positive ratings from 87.9\% (58/66).} \revision{In the multi-selection question, 63.6\% (42/66) selected ``\textit{I felt free to express myself},'' 65.2\% (43/66) selected ``\textit{I could answer at my own pace},'' and 47.0\% (31/66) selected ``\textit{I felt less pressure than in an interview with a human researcher}.''} \revision{These self-reports suggest that human absence reduced perceived pressure for some respondents, but the study did not compare their answers with those produced in a human interview.}

\revision{The comparative item produced a mixed distribution: 42.4\% (28/66) selected the midpoint, 37.9\% (25/66) leaned toward its human-interviewer endpoint, and 19.7\% (13/66) leaned toward its AI-interviewer endpoint.} \revision{Because its wording and endpoints represented partially different constructs, this distribution should be interpreted cautiously rather than as a validated preference measure.} \revision{The absence of a human researcher was rated positively by 43.9\% (29/66), at the midpoint by 45.5\% (30/66), and negatively by 10.6\% (7/66).} \revision{Thus, convenience and acceptability did not imply a uniform preference for removing human interaction.}
\revision{The most frequent reported feeling was curiosity, selected by 43.9\% (29/66), followed by comfort at 28.8\% (19/66) and interest at 15.2\% (10/66) (Figure~\ref{fig:additional-perceptions}).} Some participants also reported neutrality, strangeness, distrust, or tiredness. \revision{A total of 84.8\% (56/66) would recommend the format, 13.6\% (9/66) answered ``Maybe,'' and 1.5\% (1/66) answered ``No.''} \revision{These results reinforce acceptability among respondents while showing that the experience was not uniformly positive.}

The additional comments reinforce this interpretation. Participants described the experience as interesting, useful, and comfortable, highlighting the possibility of answering at their own pace, expressing themselves freely, and, in some cases, feeling less judged than they might in a human-conducted interview. At the same time, the comments pointed to limitations already observed in the closed-ended responses, including unnatural voice interaction, voice recognition problems, generic or repetitive questions, and limited adaptation to participants' specific contexts. These comments suggest that deeper probing, contextual adaptation, more natural interaction, and technical reliability remain important areas for improvement.

\begin{tcolorbox}[
    colback=gray!8,
    colframe=black!60,
    title={\faLightbulb\quad Finding for RQ$_{2}$},
    fonttitle=\bfseries,
    sharp corners,
    boxrule=0.6pt
]
\noindent
\revision{The analyzed respondents generally perceived the AI-conducted interview as clear, comfortable, practical, and acceptable for a short interaction. Responses were more mixed regarding human absence and the comparative item. These perceptions describe the participant experience and do not establish the richness, completeness, or accuracy of the interview evidence.}
\end{tcolorbox}

\subsection{RQ$_3$: Benefits, Limitations, and Lessons Learned}
\label{subsec:rq3-lessons}

Our third RQ investigates the benefits, limitations, and lessons learned from using AI as an interviewer in ESE. \revision{The first opportunity is logistical: the implemented workflow allowed successful completers to participate asynchronously through a link without scheduling a live researcher-led session.} \revision{We did not measure recruitment yield, researcher labor, financial cost, or net time savings, so the evidence concerns workflow design rather than quantified efficiency gains.}
\revision{Participant flexibility was also frequently selected in the multi-selection strengths question (Figure~\ref{fig:additional-perceptions}a): 65.2\% (43/66) reported answering at their own pace, 63.6\% (42/66) selected that the experience was quick, and 57.6\% (38/66) selected that it was practical and easy to complete.} \revision{These findings support perceived convenience for short interviews among the analyzed respondents.}

\revision{A common prompt can promote protocol consistency, and 65.2\% (43/66) selected that the AI kept the interview focused on the topic.} \revision{However, only 19.7\% (13/66) selected that follow-up questions helped them provide more detailed answers.} \revision{Thus, consistency should not be conflated with adaptive probing quality.}
\revision{The workflow also produced a structured artifact immediately after each completed interview, which can support initial organization.} \revision{Because we did not measure researcher effort and did not compare artifacts with full transcripts, we cannot quantify time savings or establish that the summaries preserved all relevant evidence.} \revision{Generated outputs must therefore be treated as AI-mediated artifacts rather than as verbatim transcripts.}
\revision{The most frequent limitations (Figure~\ref{fig:additional-perceptions}b) were questions perceived as too generic and insufficient sensitivity to participants' answers, each selected by 28.8\% (19/66), followed by limited depth at 27.3\% (18/66).} \revision{These limitations are central to the trade-off between protocol consistency and the contextual judgment normally exercised by a skilled human interviewer.}
\revision{Privacy concerns were selected by 22.7\% (15/66), which is important because practitioners may discuss organizations, clients, systems, products, or internal processes through an external platform.} \revision{Other reported limitations included missed human interaction at 22.7\% (15/66), repetitive questions at 19.7\% (13/66), superficiality at 13.6\% (9/66), and tool difficulty at 6.1\% (4/66).} \revision{At the same time, 19.7\% (13/66) selected that they perceived no weaknesses, showing that these concerns were not universal.}

Based on these results, we derive the following lessons learned. First, participant instructions are as important as the AI prompt. Because participants conduct the interview independently, the success of the procedure depends on whether they understand how to access the tool, use voice mode, request language changes, generate the artifact, and submit the result. \revision{In our deployment, the submission form asked participants to paste the interview ``transcript,'' while the protocol produced a structured synthesis; this terminological mismatch may explain why a few participants pasted conversation excerpts, unstructured answers, or other content instead of the synthesis.} \revision{Instructions and form wording should therefore name the expected artifact precisely, and submitted artifacts should be verified upon receipt so that participants can be asked to resubmit when the wrong content is provided.} Second, voice mode improves accessibility and naturalness, but it introduces technical risks such as transcription errors and misinterpreted sounds. Third, AI-generated summaries should be validated whenever possible, because they may omit, reorganize, or alter details. Fourth, researchers should avoid using AI-conducted interviews as the only data collection method when the study requires sensitive topics, deep emotional engagement, or detailed discussion of code, diagrams, screenshots, or other artifacts. Finally, post-interview questionnaires are valuable because they reveal how participants experienced the method, not only whether the method produced data.

\begin{tcolorbox}[
    colback=gray!8,
    colframe=black!60,
    title={\faLightbulb\quad Finding for RQ$_{3}$},
    fonttitle=\bfseries,
    sharp corners,
    boxrule=0.6pt
]
\noindent
\revision{The MyGPT-based workflow enabled asynchronous participation, a common interview structure, and immediate generation of a structured artifact among valid completers. The study did not quantify efficiency gains or assess equivalence to human interviews. Researchers must account for generic probing, limited sensitivity, insufficient depth, platform-related privacy concerns, summarization risks, and the need for human validation.}
\end{tcolorbox}

\subsection{\revision{Descriptive Content of the Submitted Interview Artifacts}}
\label{subsec:topic-insights}

\revision{Although the main goal of this article is methodological, 65 of the 66 submitted artifacts contained recognizable topic-specific content.} \revision{Two of these artifacts, however, conflicted with the protocol reported in the corresponding questionnaire record and were not reassigned without participant confirmation.} \revision{We therefore base the domain-specific descriptions below on 63 protocol-consistent artifacts: 32 associated with the refactoring protocol and 31 associated with the Scrum protocol; the unusable artifact and the two mismatched artifacts contribute only to the operational analysis.} \revision{We report this content only to illustrate what appeared in the AI-mediated artifacts, not to evaluate interview richness, construct a complete qualitative theory, or establish that the summaries faithfully represented the original conversations.} \revision{The observations below should therefore be interpreted as descriptive themes present in the protocol-consistent submitted artifacts.}

In the refactoring interviews, participants described refactoring as an opportunistic and context-dependent practice, often interleaved with feature implementation, bug fixing, code review, module stabilization, or technical debt reduction. This pattern is consistent with floss refactoring, in which refactoring is performed together with other development activities rather than as a dedicated task~\cite{Murphy-Hill-TSE-2012}, and with prior evidence that \revision{refactoring is frequently interleaved with other development tasks}~\cite{golubev-fse-2021}. Participants reported motivations such as improving readability, reducing duplication, simplifying complex logic, extracting or moving methods and classes, renaming program elements, separating responsibilities, and improving maintainability, which aligns with prior evidence on pragmatic motivations for refactoring~\cite{why-refactor-2016,Fowler-book-1999}.
A central concern in the refactoring artifacts was behavior preservation. Participants mentioned risks such as regressions, hidden dependencies, undocumented business rules, insufficient test coverage, and unintended changes in critical code. These concerns are well founded, since prior studies have shown that refactoring implementations can introduce faults or fail to preserve behavior in some cases~\cite{wang2024empiricalstudyrefactoringengine,test-tools-fse07,Soares-TSE-2013}. To mitigate these risks, participants reported relying on tests, code review, CI pipelines, QA validation, manual testing, and, in some cases, feature flags or staged deployment. The submitted artifacts frequently associated tests with confidence after refactoring. However, the risk of relying on inadequate test coverage was not prominent in these artifacts, although this limitation has been documented in prior work~\cite{DBLP:conf/icsm/RachatasumritK12}. Because the artifacts may compress or omit details from the original conversations, we cannot determine whether participants discussed this issue during the interviews. Participants also reported using tool support, including IDEs, static analysis tools, CI pipelines, version-control workflows, and AI-based assistants, consistent with prior evidence on refactoring tool usage~\cite{oliveira2019revisiting}. However, some of them viewed these tools as useful but insufficient without human review, especially for complex or high-risk refactorings, for which developers often prefer to apply changes manually rather than relying fully on automated refactoring tools~\cite{Tempero-ACM-2017}.

In the Scrum and generative AI interviews, participants described AI mainly as a support tool for writing, organizing, summarizing, and preparing work artifacts. Reported uses included product backlog refinement, user story writing, acceptance criteria generation, ticket creation, documentation, release notes, meeting summaries, retrospectives, sprint planning support, technical task decomposition, estimation support, test-related activities, and preparation for sprint reviews. Some participant comments also framed generative AI in Scrum as a topic that requires careful reflection, especially regarding productivity, team maturity, validation practices, responsible use, and its effects on work dynamics. Participants generally framed AI as an accelerator rather than a decision maker: AI reduced operational effort and improved the initial structure of artifacts, but priorities, business rules, stakeholder alignment, compliance, and team commitments remained human accountabilities. Adoption was uneven across participants and teams, ranging from daily or recurrent use with tools such as ChatGPT, Copilot, Gemini, Claude, Jira integrations, spreadsheets, IDEs, or internal assistants to experimental, individual, or informal use. Reported risks included generic or incorrect outputs, hallucinated or incomplete acceptance criteria, missing product or architectural context, overreliance, validation effort, and confidentiality concerns.
These findings are consistent with recent work on AI and LLMs in agile software development. Cinkusz and Chudziak~\cite{cinkusz2024towards} discuss the potential of LLM-augmented multi-agent systems to support agile software engineering activities, particularly coordination, communication, and process support. Perkusich et al.~\cite{perkusich2026adoption} provide empirical evidence from Scrum practitioners, reporting that LLMs support knowledge-intensive Scrum activities and are associated with perceived benefits such as productivity gains and reduced manual effort, while still raising risks related to almost-correct outputs, hallucinations, confidentiality, trust, and organizational readiness. Recent interview-based evidence on GenAI adoption in agile teams also reports similar patterns, including use for documentation and creative tasks, perceived efficiency gains, and barriers related to privacy, validation effort, and governance~\cite{neumann2026between}. \revision{The thematic overlap indicates that the submitted summaries contained domain-relevant material, but it does not by itself demonstrate depth, completeness, or fidelity.}

Across both topics, the artifacts suggest that participants valued automation when it reduced effort, accelerated routine work, or improved organization, but they also emphasized that human validation remained essential. In refactoring, validation focused on tests, code review, behavior preservation, regression control, and the interpretation of risky code changes. In Scrum-related AI use, validation focused on domain correctness, business rules, confidentiality, stakeholder alignment, and fit with the team's process. \revision{These recurring themes show that the submitted artifacts were topically relevant, but the absence of full transcripts prevents conclusions about what the AI failed to probe, omitted, or reformulated.} \revision{The artifacts may support preliminary descriptive organization and identification of follow-up topics, but stronger qualitative claims require fuller evidence and human-led validation.}

\section{Threats to Validity}
\label{sec:threats}

\textbf{Construct validity.}
\revision{The questionnaire measures participant perceptions---including comfort, clarity, naturalness, comparative orientation, and acceptability---but it does not directly assess the depth, richness, completeness, or accuracy of the interview evidence.} \revision{Moreover, the comparative item combined wording about how participants felt they answered with endpoints framed as interviewer preference, creating ambiguity in the measured construct; we therefore interpret it cautiously.}
\revision{The submitted artifacts are AI-generated syntheses rather than standardized full transcripts. The AI may summarize, compress, reorganize, omit, or rephrase participant statements, and the procedure did not include systematic member checking.} \revision{Inspecting the artifacts for expected structure and topic-relevant content mitigates obvious incompleteness but cannot validate fidelity to the original conversation; indeed, this inspection revealed that a small number of submissions did not contain the expected synthesis or contained unusable content (Section~\ref{subsec:rq1-feasibility}), confirming that artifact verification is a necessary step in this kind of workflow.} \revision{The two protocol--artifact mismatches further show that self-administered submission can introduce linkage errors between questionnaire metadata and the material pasted by participants; without participant confirmation, such records should not be silently reassigned for domain-specific analysis.} \revision{The structural audit was conducted by one researcher, with ambiguous and protocol-inconsistent cases discussed by the research team; because the classifications were not independently replicated by a second coder, they remain subject to researcher judgment.}

\textbf{Internal validity.}
Participants completed the procedure independently, using their own devices, network connections, browsers, ChatGPT accounts, and available models. Differences in audio quality, familiarity with voice interaction, language handling, and ability to follow instructions may have affected the interview experience. \revision{Free and paid accounts may also have differed in model access, limits, or interface behavior, and the exact model could not be standardized across interviews.} \revision{These factors may have influenced language switching, follow-up questions, response quality, and synthesis generation.} \revision{We mitigated this threat through step-by-step instructions, researcher-led pilot testing, access checks using free accounts, and documentation of observed problems, but these measures do not ensure identical interviewer behavior across participants.}
The AI interviewer itself is also a source of variation. Even with the same prompt, an LLM-based interviewer may ask different follow-up questions, vary in probing depth, or handle language changes inconsistently. Moreover, model updates, interface changes, account-specific access conditions, and different ways of requesting the final structured synthesis may affect the generated artifact and make exact replication difficult. To support reliability, we preserve the interview prompts, participant instructions, questionnaire instrument, and analysis artifacts in the replication package.
\revision{Participants' prior familiarity with AI tools, voice mode, and research interviews may have influenced their perceptions, especially in the Scrum study.} \revision{These characteristics were not measured in sufficient detail for subgroup analysis.} The novelty effect may also have made the experience seem more convenient, engaging, or unusual than it would be after repeated use. To mitigate this threat, the questionnaire included both positive and negative items, allowing participants to report strengths, limitations, discomfort, strangeness, preference for human interaction, and technical difficulties.

\textbf{External validity.}
\revision{Our findings are based on \participants{} completed submissions from participants living in one country and across two short, technology-oriented ESE studies.} \revision{The submitted artifacts were written almost exclusively in Portuguese (65/66), and interaction language was not recorded independently; consequently, the multilingual capability of the workflow, although piloted in three languages, was barely evidenced in practice and the results provide limited evidence about cross-language use.} \revision{Because the analysis does not use a complete denominator for invitations or incomplete attempts, selection and completion bias cannot be quantified.} \revision{The breadth obtained from two topics also limits depth and makes it difficult to separate effects of the interviewer, topic, protocol, and participant background.} \revision{The results may not generalize to sensitive topics, emotionally complex experiences, conflict-heavy workplace issues, long interviews, artifact-rich sessions, or studies requiring deep rapport and extensive contextual probing.}
\revision{In addition, the implementation is tightly coupled to OpenAI's MyGPT infrastructure: several observations---including account-tier differences, interface and voice-mode behavior, and language switching---are platform-specific and may not generalize to other LLM ecosystems or to future versions of ChatGPT.} \revision{For this reason, we treat the prompts, instructions, questionnaire, and workflow---rather than the specific platform behavior---as the primary transferable artifacts.}
\revision{The \participants{} submissions were received within three days, showing that the workflow can collect completed responses without scheduling synchronous sessions.} \revision{This does not quantify recruitment effectiveness or establish that the resulting data are equivalent in quality to human-conducted interviews.} The interviews lasted approximately 10--15 minutes and were designed to collect focused reflections. \revision{Our findings should therefore be interpreted as evidence of operational viability and perceived acceptability among completers of short interviews.} Our voice-based workflow was also less suitable for artifact-rich interviews involving code snippets, pull requests, diagrams, screenshots, issue reports, or architectural documents.

\textbf{Conclusion validity.}
Because this is an experience report with an exploratory design, our conclusions are descriptive rather than causal. \revision{We do not claim that AI-conducted interviews produce data of equal or superior quality to human-conducted interviews, that the submitted summaries faithfully represent the original conversations, or that positive perceptions would hold in other settings or among non-completers.} \revision{The evidence instead indicates a trade-off: this implementation enabled asynchronous participation and natural-language flexibility, while limiting researcher control, rapport, probing depth, and confidence in artifact fidelity.}

\section{Related Work}
\label{sec:related-work}

Interviews are an established method in ESE. Hove and Anda~\cite{hove2005experiences} synthesize experiences from 12 ESE studies involving 280 interviews and highlight challenges related to effort estimation, interviewer skills, interaction quality, and the use of appropriate tools and project artifacts during interviews. Nasar~\cite{nasar2023expertinterviews} also discusses expert interviews in software engineering, emphasizing their value for eliciting domain knowledge while noting challenges related to access to experts, biased information, terminology differences, language use, and ethical concerns. These studies are important for our work because they clarify why interviews are valuable but costly: human interviewers can adapt the conversation, ask clarifying questions, build rapport, and decide when deeper probing is needed, but interview studies also require substantial coordination, scheduling, transcription, and analysis effort.

Ethical concerns are also central to interview-based software engineering research. Strandberg~\cite{DBLP:conf/esem/Strandberg19} proposes ethical guidelines and checklists for software engineering interview studies, emphasizing consent, confidentiality, anonymization, scientific value, researcher skill, legal compliance, and ethical review. Our study builds on these concerns in a new setting: self-administered interviews mediated by an external AI platform. This setting preserves traditional ethical concerns while adding new ones, such as participant control over shared artifacts, risks of disclosing confidential information to the AI system, and the need to treat AI-generated summaries as mediated research artifacts rather than verbatim transcripts.

Recent work has explored how AI and LLM-based systems can support qualitative data collection. Chopra and Haaland~\cite{chopra2023conducting} investigate AI-conducted qualitative interviews at scale and suggest that AI interviewers can reduce cost while maintaining participant satisfaction. Cuevas et al.~\cite{cuevas2023qualitative} evaluate LLM-based chatbots for collecting qualitative data and show that such systems can elicit useful responses, while also struggling to capture participants' specific motives and personalized contextual examples. Zhang et al.~\cite{zhang2025harnessing} study LLM-generated follow-up questions in semi-structured interviews, showing that AI can support interviewers but still requires human judgment and oversight. These studies are closely related to ours because they examine AI-supported qualitative interviewing; our work complements them by studying a self-administered AI-interview workflow in ESE, including participant perceptions and the methodological implications of relying on AI-generated interview artifacts.

\revision{Related work has also investigated autonomous conversational data collection and AI support for specific interviewing skills. Xiao et al.~\cite{xiao2020tell} compared an AI-powered conversational survey with a conventional online survey and found that the conversational format increased participant engagement and elicited responses that were more informative, relevant, specific, and clear. In subsequent work, Xiao et al.~\cite{xiao2020hear} designed interview chatbots with active-listening capabilities, demonstrating the importance of interpreting participants' free-text answers and responding in ways that acknowledge and build on their input. Hu et al.~\cite{hu2024designing} examined how conversational agents can generate follow-up questions for information elicitation, a capability that is central to the depth and adaptability of semi-structured interviews. Liu et al.~\cite{liu2025envisioning} investigated researchers' expectations of AI assistance across different levels of interviewing expertise, finding support for designs in which AI assists the interviewer without replacing human judgment or taking control of the interaction. These studies highlight a spectrum ranging from autonomous conversational data collection to human-in-the-loop interviewing support. Our work examines the autonomous end of this spectrum, in which the AI conducts the interaction without a researcher present, while InterFlow~\cite{wen2026interflow} represents a hybrid approach that preserves human control over probing, clarification, and interview flow.}

Gerosa et al.~\cite{gerosa2023ai} discuss whether AI can serve as a substitute for human subjects in software engineering research; in contrast, we preserve human participants as the source of empirical evidence and use AI as the interviewer. Prior work in requirements engineering has explored LLMs for interview-related workflows, such as simulating stakeholders, supporting requirements elicitation practice, generating requirements from interview data, or assisting with transcript analysis~\cite{lojo2025requirements,almeida2025requirements}; however, these studies focus primarily on requirements outcomes rather than the interview method itself. Mejias et al.~\cite{mejias2023chatgpt} and Rusu et al.~\cite{rusu2011simulating} explore computational support for technical interview preparation, using ChatGPT as a mock interviewer or a serious game to simulate interview scenarios. These works show that computational systems can support interview-like interactions, but they target training and assessment rather than qualitative data collection. In contrast, our study evaluates self-administered, voice-based AI interviews in real ESE studies, focusing on feasibility, participant perceptions, and methodological implications.

Overall, there is still limited evidence about how software engineering participants experience self-administered AI-conducted interviews in real empirical studies, particularly when interviews are conducted through voice interaction rather than text-only chat. \revision{Our work addresses this gap as an experience report on the design, deployment, participant perceptions, and limitations of a complete self-administered workflow in two ESE contexts.} \revision{Unlike a controlled comparison of interview methods, our contribution concerns operational experience and acceptability; the quality and fidelity of AI-mediated artifacts remain open empirical questions.}

\section{Conclusions}
\label{sec:conclusion}

\revision{This experience report documented the design and deployment of a MyGPT-based, self-administered, voice-oriented interview workflow in two ESE studies and analyzed \participants{} submissions. The evidence concerns operational execution and participant perceptions among completers, rather than methodological equivalence between AI and human interviewing. Respondents generally rated the short interaction positively, but the comparative item and reported weaknesses showed more mixed views about human absence, probing depth, privacy, and interaction quality. The artifact audit found that 61/66 submissions followed the expected structured-synthesis format, while five did not, and it identified two protocol--artifact inconsistencies. 

These results show that acceptability and structural completeness do not establish the richness, fidelity, or analytical value of the underlying qualitative evidence.
The workflow may therefore be useful as a complementary option for short, focused, low-risk, and asynchronous exploratory data collection, particularly when a live session is difficult to schedule. Researchers adopting similar workflows should document platform conditions, pilot the complete participant procedure, provide explicit privacy guidance, name the expected artifact precisely, verify submissions upon receipt, collect full transcripts when possible, and validate AI-generated syntheses before making substantive qualitative claims.}

\revision{Future controlled studies should compare AI-only, human-only, and hybrid interviews using the same protocol and comparable participants, assessing probing behavior, clarification, richness, completeness, fidelity, coding outcomes, participant experience, and researcher effort. Further work should also examine focused single-domain studies, variation across models and platforms, natural-language and voice conditions, interview length, and artifact-rich settings.}


\end{document}